\documentclass[useAMS,usenatbib,usegraphicx]{mn2e}
\usepackage{aas_macros}
\usepackage{amssymb}

\title[The Star Formation History in The Far Outer Disc of M33]
{The Star Formation History in The Far Outer Disc of M33}
\author[Barker et al.]
{{\upshape \large Michael K. Barker$^1$\thanks{mkb@roe.ac.uk}, 
A. M. N. Ferguson$^1$\thanks{ferguson@roe.ac.uk},
A. A. Cole$^2$, R. Ibata$^3$, M. Irwin$^4$,}\\
{\upshape \large G. F. Lewis$^5$, T. A. Smecker-Hane$^6$, N. R. Tanvir$^7$}\\
$^1$Institute for Astronomy, University of Edinburgh, Blackford Hill, 
Edinburgh, UK, EH9 3HJ,\\
$^2$School of Mathematics \& Physics, University of Tasmania, Private Bag 37, Hobart, 7001, TAS, Australia,\\
$^3$Observatoire Astronomique de Strasbourg, Strasbourg, France,\\
$^4$Institute of Astronomy, Cambridge University, Cambridge, UK,\\
$^5$Sydney Institute of Astronomy, School of Physics, The University of Sydney, NSW 2006, Australia\\
$^6$Department of Physics \& Astronomy, University of California, Irvine, USA\\
$^7$Department of Physics and Astronomy, University of Leicester, Leicester, UK
}

\begin{document}

\newcommand{\MITRGB}[0]{{\rm M_{I}(TRGB)}}
\newcommand{\vmi}[0]{\rm (F606W-F814W)}
\newcommand{\msun}[0]{\rm M_{\sun}}
\newcommand{\msunyr}[0]{\rm M_{\sun}\ yr^{-1}}
\newcommand{\msunpc}[0]{\rm M_{\sun}\ pc^{-2}}


\date{Accepted ----. Received ----; in
  original form ----}

\pagerange{\pageref{firstpage}--\pageref{lastpage}} \pubyear{2010}

\defcitealias{Eggen62}{ELS62}
\defcitealias{Searle78}{SZ78}
\defcitealias{Pagel95}{PT95}
\defcitealias{Pagel98}{PT98}
\defcitealias{Barker07a}{Paper II}
\defcitealias{Barker07b}{B07}
\defcitealias{Williams09b}{W09b}
\defcitealias{Roskar08a}{R08a}
\defcitealias{Roskar08b}{R08b}

\maketitle

\label{firstpage}

\begin{abstract} 

The outer regions of disc galaxies are becoming increasingly
recognized as key testing sites for models of disc assembly
and evolution.
Important issues are the epoch at which the bulk of the stars
in these regions formed and how discs grow radially over time.
To address these issues, we use 
{\it Hubble Space Telescope} Advanced Camera for Surveys
imaging to study the star formation history (SFH) of two fields at 9.1 
and 11.6 kpc along M33's northern major axis.
These fields lie at $\sim 4$ and 5 $V$-band disc scale-lengths
and straddle the break in M33's surface brightness profile.
The colour-magnitude diagrams (CMDs) reach 
the ancient main sequence turnoff 
with a signal-to-noise ratio of $\sim 5$.
From detailed modelling of the CMDs,
we find that the majority of stars in both fields combined
formed at $z < 1$.
The mean age in the inner field, S1, is $\sim 3 \pm 1$ Gyr
and the mean metallicity is [M/H] $\sim -0.5 \pm 0.2$ dex.
The star formation history of S1 unambiguously reveals how the inside-out
growth previously measured for M33's inner disc out to $\sim 6$ kpc
extends out to the disc edge at $\sim 9$ kpc.
In comparison, the outer field, S2, 
is older (mean age $\sim 7 \pm 2$ Gyr), 
more metal-poor (mean [M/H] $\sim -0.8 \pm 0.3$ dex),
and contains $\sim 30$ times less stellar mass.
These results provide the most compelling evidence yet that M33's
age gradient reverses at large radii near the disc break
and that this reversal is accompanied by a break in
stellar mass surface density.
We discuss several possible interpretations of this behaviour
including radial stellar mixing, 
warping of the gaseous disc, 
a change in star formation efficiency, and 
a transition to another
structural component.
These results offer one of the most detailed views yet 
of the peripheral regions of any disc galaxy 
and provide a much-needed observational constraint
on the last major epoch of star formation in the outer disc.

\end{abstract}


\section{Introduction}
\label{sec:intro}

The traditional theory of disc galaxy formation holds that 
isolated discs form through the dissipational
collapse of gaseous protogalaxies
embedded in cold dark matter (CDM) haloes \citep{White78,Fall80,Peebles84}.
N-body/SPH simulations of structure formation in a CDM Universe have shown 
the importance of the cosmological context to disc formation, as
galaxies grow through the hierarchical
merging and accretion of many smaller systems and through the inflow 
of intergalactic gas \citep[e.g.][]{Steinmetz02,Abadi03,
SommerLarsen03,Governato09}.
The relative proportion of these two growth mechanisms
through time may have a large effect on a galaxy's morphology today, 
as gas accretion tends to build thin discs and
merging tends to build spheroids and thick discs
\citep{Steinmetz02,Brook04,Brooks09}.
Other works have stressed the importance of 
gas-rich major mergers as an alternate or complementary
disc formation channel to the ones above
\citep{Robertson06,Stewart09}.
In reality, all of these mechanisms may operate
at different levels depending on a galaxy's environment
\citep{Blanton03}
and may contribute to diverse properties of outer discs
and haloes.

Simulations of disc galaxies in a cosmological context
have advanced to the point where they can make detailed predictions for
their assembly history and observable internal properties.
A common prediction is that present-day thin discs formed the bulk
of their stars relatively late, at $z \lesssim 1$,
because the major merger rate at earlier epochs was too high
for any pre-existing thin disc to survive
\citep{Abadi03,SommerLarsen03,Governato09}.
While there is still uncertainty in their
sub-grid physical models of star formation, stellar feedback, 
and the multi-phase ISM \citep[e.g.][]{Okamoto05}, the simulations agree that 
stellar ages and chemical compositions contain clues to the
formation and evolution of spiral galaxies.
Therefore, spatially-resolved observational estimates
of the star formation histories (SFHs) within spirals can provide
powerful constraints on the simulations and
their sub-grid physical models.
These constraints are more robust against degeneracies
when they come from colour-magnitude diagrams (CMDs) 
of resolved stars than when they rely on integrated, unresolved 
starlight.  Some progress in this area has been made 
using the Advanced Camera for Surveys (ACS) 
aboard the {\it Hubble Space Telescope} to observe galaxies 
in the Local Group (LG) and nearby Universe 
\citep{Brown06,Williams09a}. 
However, the small field-of-view of ACS
often makes it unclear which galactic component, or mix 
of components has been imaged (thin disc, thick disc, bulge, halo, 
accreted substructure, disturbed disc, etc.).  
This problem is particularly acute in MW-type 
spirals, which have non-negligible bulges and are more 
likely to have experienced massive satellite accretion, 
which can hinder efforts to study purely dissipative disc formation
and secular evolution.

The late-type LG spiral, M33, potentially 
offers a clearer view of disc evolution than M31 or 
the MW because it has little or no bulge component 
\citep{Bothun92,Minniti93,Mclean96}.  
Such bulgeless disc galaxies are common in the Universe and the most 
challenging to produce in cosmological simulations, 
so a better observational determination of their 
evolution provides a rigorous benchmark for our 
understanding of disc formation \citep{Mayer08,Kautsch09}.
M33 is also interesting because it is one of a class of 
galaxies that exhibit truncations in their radial 
surface brightness (SB) profiles \citep{Pohlen06,Ferguson07a}.  
What causes these truncations is unclear, but they
could arise naturally from the collapse of a uniformly-rotating,
homogeneous protogalactic gas cloud under angular momentum
conservation \citep{vanderkruit87}, or they could be related to 
star formation thresholds \citep{Elmegreen94,vandenBosch01,Schaye04}.
In M33, the break radius in the azimuthally averaged SB
profile lies at $\sim 36\arcmin$ \citep{Ferguson07a} and
the $V$-band inner disc scale-length is $\sim 9.2\arcmin$
\citep{Guidoni81}.  The optical radius is
$R_{25} = 35.4\arcmin$, defined as the radius at which the B-band
SB is $\rm 25.0\ mag\ arcsec^{-2}$ \citep{deVauc91}.

Using CMDs of resolved stars in M33, 
\citet[][hereafter B07]{Barker07b} found a 
positive age gradient in three ACS fields just 
outside the break radius on the southern minor axis.  
More recently, \citet[][hereafter W09b]{Williams09b} 
found a negative age gradient in four ACS fields 
within the break radius on M33's southern major axis.  
\citetalias{Williams09b} argued that these seemingly contradictory results 
could be reconciled by the simulation of 
\citet{Roskar08b}, in 
which the inner disc of a spiral galaxy forms inside-out, 
but secular processes radially redistribute stars 
causing an inflection point in the stellar age 
gradient at the break radius.

Some open questions remain, however, since the 
B07 and \citetalias{Williams09b} fields sample a small fraction of M33 
and lie at different position angles.  As shown 
by \citet{McConnachie06} from spectroscopy of 
red giant branch (RGB) stars and by \citet{Sarajedini06} from the 
periods of RR Lyrae stars, M33 hosts a metal-poor 
field halo population.  \citet{McConnachie06} 
also found 
three kinematic components on the southern major axis
including a dominant disc, minority halo, and a
third component which they attributed to a possible
tidal stream.
More recently, 
\citet{McConnachie09} uncovered two giant tidal
tails extending from the northern and southern
disc regions.  Hence, it is important to 
measure the SFH at as many different locations as 
possible to better understand and disentangle the 
contributions of different structural components 
and to safeguard against spatial variations 
within any one component.

\begin{figure}
\includegraphics[width=80mm,height=150mm,keepaspectratio=true]{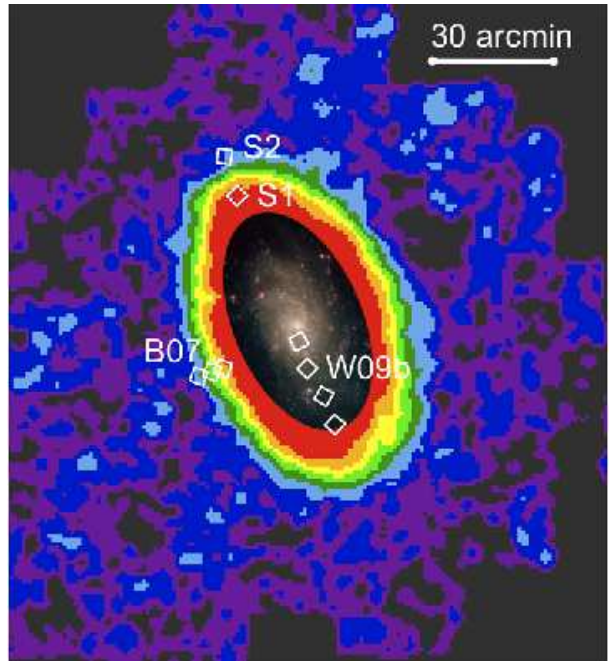}
\caption{
Contour map of M33 RGB stars from the INT/WFC survey with an
optical image of the inner disc overlaid.
The coloured contours represent equal logarithmic intervals
of RGB stellar surface density.
The ACS fields studied in this paper are labelled S1 and S2.
The fields studied by \citetalias{Barker07b} and \citetalias{Williams09b} 
lie on the southern minor and major axes, respectively.
The legend in the upper right corner shows the scale of 30\arcmin.
}
\label{fig:fields}
\end{figure}

In this paper, we examine the stellar populations 
in two ACS fields located at $36\arcmin$ and $46\arcmin$ 
along M33's northern major axis.
The fields lie at $\sim 4$ and 5 $V$-band scale-lengths, 
or 9.1 and 11.6 kpc, 
assuming a distance of 867 kpc \citep{Galleti04}, 
inclination of $56\degr$ and position angle of $23\degr$ \citep{Corbelli97}.
Fig. \ref{fig:fields} shows the locations of these 
fields, and those studied by \citetalias{Barker07b} 
and \citetalias{Williams09b}, on a contour map of 
RGB stars from the Isaac Newton Telescope 
Wide Field Camera (INT/WFC) survey
of M33 \citep{Ferguson07a}.
The contours represent equal logarithmic intervals
of RGB stellar surface density.
Overlaid on the RGB map is an optical image of the inner disc on the same 
spatial scale.
Our two fields, henceforth called S1 and S2, straddle the break in 
M33's SB profile and provide one of the deepest views 
yet of the stellar populations in the outskirts of a disc galaxy.

In the next section, we describe the observations
and data reduction and 
in \S \ref{sec:cmds}, we examine the CMDs.
We outline in \S \ref{sec:sfhmeth} our method for 
fitting the CMDs to obtain the SFH.
In \S \ref{sec:sfhres}, we present the results
of the CMD fitting analysis and
in \S \ref{sec:disc}, we discuss the implications.
Finally, we summarise our results in \S \ref{sec:conc}.

\section{Observations}
\label{sec:obs}

The observations of S1 and S2 were made 
in June 2004 and August 2004, respectively,
for Program ID 9837 (PI: Ferguson).
The coordinates were $(\alpha_{J2000}, \delta_{J2000}) = 
(01^h34^m58.8^s, +31^\circ12'00.0'')$ for S1 and
$(01^h35^m15.0^s, +31^\circ22'00.0'')$ for S2.
Each field was observed over a period of
12 orbits using the F606W (broad $V_J$) and F814W ($\sim 
I_c$) filters.
To facilitate removal of cosmic rays and bad pixels, 8 dithered
sub-exposures totaling 10350s were taken in F606W and
16 sub-exposures totaling 20700s were taken in F814W.
We photometered the images using the ACS module of the 
DOLPHOT software package\footnote{DOLPHOT is an adaptation of the photometry
package HSTphot \citep{Dolphin00}. It can be downloaded
from http://purcell.as.arizona.edu/dolphot/.}.  
DOLPHOT takes as input the individual
flat-fielded, dark-subtracted FLT images produced by the CALACS
pipeline in addition to a reference image used for a common
pixel coordinate system.
We selected
the deepest F814W DRZ image automatically produced by the pipeline
to use as a reference, which was
corrected for geometric distortion.
We used the recommended values of input parameters
listed in the DOLPHOT manual except we turned off the option
to do a second pass of star finding and photometry
because this increased the scatter in the CMD.
DOLPHOT performs PSF-fitting photometry using a 
library of precomputed TinyTim PSFs, 
applies charge transfer efficiency corrections \citep{Riess04},
and reports final magnitudes in the VEGAMAG and 
Johnson-Cousins system using the zero-points and
transformations in \citet{Sirianni05}.

\begin{figure*}
\includegraphics[width=\textwidth,keepaspectratio=true]{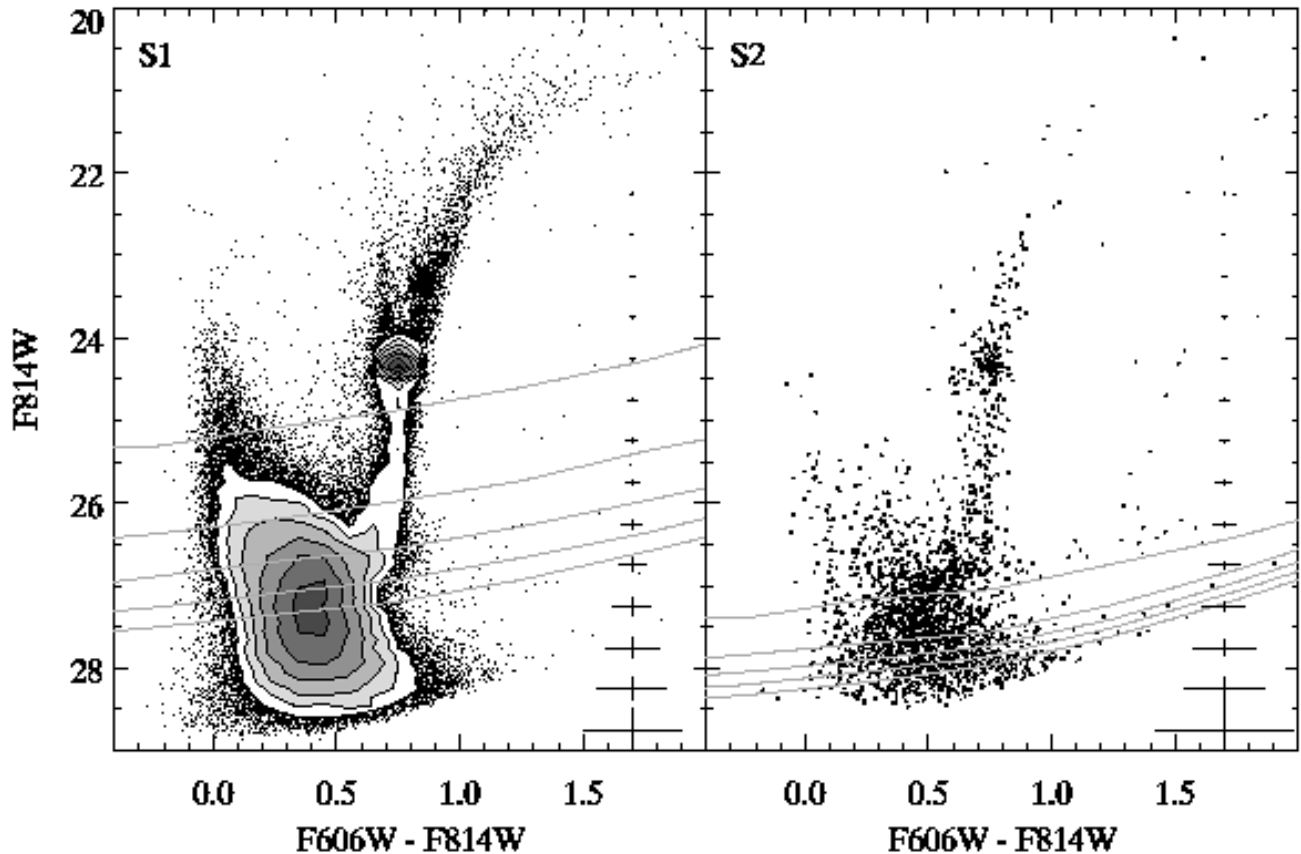}
\caption{CMD of field S1 (left) at 9.1 kpc and S2 (right) at 11.6 kpc.
The contour levels are $[1.0,1.5,2.25,3.4,5.1,7.65,11.5] \times 10^{4}
\rm mag^{-2}$.  The error bars on the right side of each panel show
median errors derived from artificial star tests.
The solid gray lines from top to bottom denote the 
90, 80, 70, 60, and 50\% completeness levels.
}
\label{fig:cmds}
\end{figure*}

To minimize contamination from non-stellar and 
poorly-measured objects in 
the final photometric catalogs, we required 
sources to have signal-to-noise ratio $> 5$ in both filters, 
$\vert F606W_{sharp} + F814W_{sharp}\vert <= 0.3$, 
$(F606W_{crowd}+F814W_{crowd}) \leq 0.1$, 
error flag $\le 7$ in both filters, and overall 
object type $\leq 2$.
The sharpness parameter measures the sharpness of
sources; good stars have values close to zero
while extended objects and incompletely masked cosmic rays
tend to have large negative and positive values, respectively.
The crowding parameter tells by how many magnitudes
brighter a star would have been recovered had nearby
stars not been fit simultaneously.
Large values suggest that a star's magnitude
was seriously affected by neighbors.
We settled on thresholds for these parameters that
produced clean CMDs and the best possible fits to the CMDs.
Additionally, an error flag $> 7$ indicates an extreme case
of saturation or that too much of the photometry aperture 
extends off a chip edge,
and objects with type $> 2$ are too sharp, extended, or elongated.
Fig.\ \ref{fig:cmds} shows the CMDs for the final
catalogs, which contain 72068 and 2304 sources in S1 and S2, respectively.
The contours in the left-hand panel of Fig.\ \ref{fig:cmds}
correspond to stellar densities of
$[1.0,1.5,2.25,3.4,5.1,7.65,11.5] \times 10^{4}\ \rm mag^{-2}$.
The drastic difference in stellar density between the fields
despite their being located only $\sim 1$ V-band scale-length apart
is due to the fact that they bracket M33's break radius.

Artificial star tests were performed in the standard
way to assess photometric errors and completeness.
Roughly $3.5 \times 10^6$ artificial stars were injected 
uniformly on the images with 
magnitudes and colours populating the entire CMD
and they were measured in exactly the same way as
the real stars.
The error bars on the right side of each panel in Fig.\ \ref{fig:cmds} 
show the median photometric errors (i.e., the difference
between input and recovered colours and magnitudes) 
derived from the artificial star tests.
The median $F814W$ error reaches 0.1 mag at $F814W \sim 27.1$
(27.8) in S1 (S2)
and the median colour error reaches 0.1 mag at $F814W \sim 27.6$
(27.4) in S1 (S2).
The colour error reaches 0.1 mag at a fainter 
magnitude than the $F814W$ error in S1 because the errors
in the two filters are strongly correlated.
This correlation is insignificant in S2 because S2 has a much
lower level of crowding.
The completeness curves are plotted on the CMDs in Fig.\ \ref{fig:cmds}.
The artificial star tests indicate that 50\% completeness occurs
at $F814W \sim 27.3$ in S1 and 
$F814W \sim 28.0$ in S2.
To avoid large incompleteness corrections, we focus most of
our analysis on the common CMD region where both fields
have $\gtrsim 60\%$ completeness.
We note that the faintest sources measured in S2
are about 0.5 mag brighter than in S1 despite identical exposure times.
This difference is due to a much smaller angle between
the Sun and the telescope's optical axis during the S2
observations ($55\degr$ for S2 vs.\ $110\degr$ for S1), 
resulting in a sky background about 2 times brighter in S2 than in S1.
Finally, we expect the number of foreground stars to be
small in both fields.  
The Besancon model of the Milky Way \citep{Robin03} predicts
only $\sim 15$ foreground stars in the region defined
by $0 < (F606W-F814W) < 1$ and $25 < F814W < 29$.

\section{Colour-Magnitude Diagrams}
\label{sec:cmds}

\begin{figure*}
\includegraphics[width=\textwidth,keepaspectratio=true]{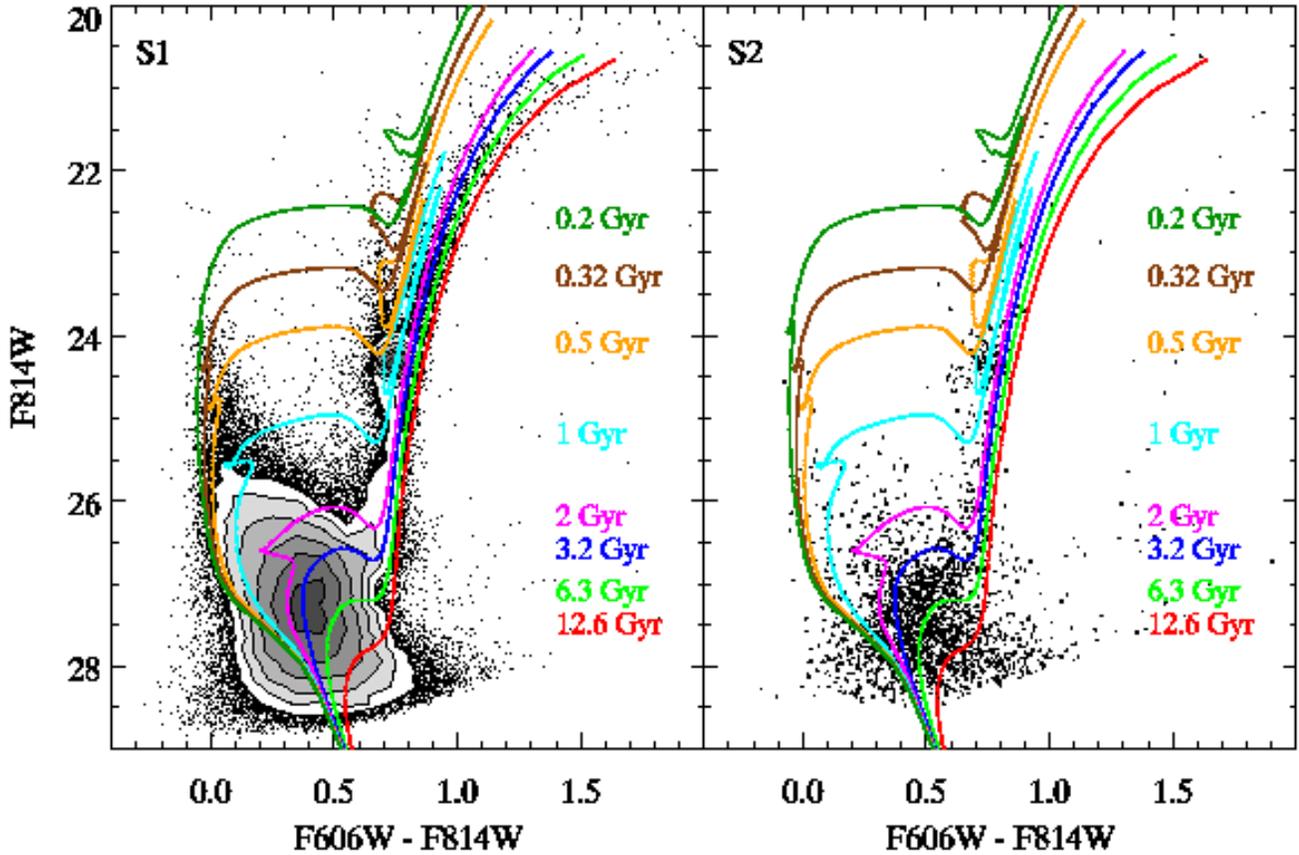}
\caption{CMDs with BaSTI theoretical
isochrones overplotted.  
The isochrones have [M/H] $= -0.7$ dex and age = 0.2, 0.32, 0.5, 1, 2, 
3.2, 6.3, and 12.6 Gyr from top to bottom.
The contours are the same as in Fig. \ref{fig:cmds}.
The magnitude of the SGB in both CMDs indicates
a significant population $\lesssim 6$ Gyr old.
}
\label{fig:cmdiso}
\end{figure*}

We begin with a qualitative inspection of the CMDs,
to be followed by a more detailed derivation of the SFH in \S \ref{sec:sfh}.
Fig.\ \ref{fig:cmdiso} shows the BaSTI theoretical isochrones 
\citep{Pietrinferni04} overplotted on the CMDs
adopting a distance modulus of $(m-M)_0 = 24.69$ 
\citep{Galleti04}.
For the reddening, we adopt the \citet{Schlegel98} 
maps and the ACS filter extinction ratios 
for a G2V star \citep{Sirianni05}.
The average values for S1 and S2 are
$E(F606W-F814W) = 0.04$ and $A_{F814} = 0.08$,
which are within 0.01 mag of the values for each field
individually.
The isochrones have a 
metallicity of [M/H] $= -0.7$ dex and ages of 0.2, 0.32, 0.5, 1.0, 2.0, 
3.2, 6.3, and 12.6 Gyr.
We use [M/H] $= -0.7$ dex as a fiducial value for both fields 
in Fig. \ref{fig:cmdiso} because it roughly matches most of the CMD features
and provides a useful starting point for a
comparative discussion.
Throughout this paper we use the common
approximation for metallicity, 
$[M/H] \approx log(Z/Z_{\sun})$ where $Z_{\sun} = 0.019$.

The sub-giant branch (SGB) is one of the most age-sensitive
and well-understood stellar evolutionary phases.
Field S1 has a prominent SGB which separates
from the main sequence (MS) at $F814W \sim 26.75$
while the SGB in S2 is $\sim 0.25$ mag fainter.
Visual inspection of the shape and position of the SGB
suggests a significant fraction of stars $< 6$ Gyr old in both fields,
consistent with several other properties of the CMDs.
For example, the mean de-reddened colour 
and magnitude of the S1 red clump (RC)
in the Johnson-Cousins system are $(V-I)_0 = 0.92 \pm 0.01$ 
and $M_I = -0.44 \pm 0.02$.
These values differ in S2 by no more than 0.03 mag.
According to the empirical calibration of red horizontal branch stars
made by \citet{Chen09}, 
the RC colour and magnitude 
give an age of $\sim 4 - 6$ Gyr
assuming [Fe/H] $\sim -0.6$ dex, 
although it should be noted that their
calibrating open cluster sample contained 
no clusters with ages less than 8 Gyr
and only one with [Fe/H] $> -0.7$ dex.
According to the models of \citet{Girardi01}, 
the RC magnitude is consistent with 
ages $\lesssim 5$ Gyr for $[M/H] > -1.7$ dex
and an age of $\sim 3$ Gyr for $[M/H] = -0.7$ dex.
Similarly, the colour is consistent with
ages $\sim 1 - 6$ Gyr and 
[M/H] $\sim -0.8$ to $-0.4$ dex.
The stellar models predict the RC magnitude to fade with increasing age
and metallicity for ages $\lesssim 10$ Gyr.
The colour distribution of RGB stars in S2 is marginally
bluer than in S1, with a K-S test giving a probability of
$6\%$ that they are drawn from the same parent distribution.
Hence, if S2 is older than S1, as indicated 
by the SGB, then it would also
have to be more metal-poor to have nearly the same RC magnitude
and RGB colour.

The number of asymptotic giant branch (AGB) stars 
above the red giant branch (RGB) tip is another
age indicator because their lifetime increases with decreasing mass
\citep{MartinezDelgado99}.
\citet{Elston92} estimated the ratio of the number of
AGB stars within one magnitude above the RGB tip to the
number of first ascent RGB stars within one magnitude below
the tip to be 0.17 -- 0.25 in the presence of a significant population
with an age of 2 -- 4 Gyr.
Although there are
too few stars to calculate this ratio in S2, we find a
value of $0.31 \pm 0.15 \rm (random)$ in S1, again suggestive of 
a large intermediate-age population.

There is no obvious horizontal branch (HB) of an old, metal-poor 
population in the CMDs of S1 or S2.
However, this feature could be
lost amongst the MS stars
and bright SGB stars crossing the Hertzprung Gap.
The presence of RR Lyrae (RRL) variable stars would unequivocally signal 
an ancient ($\gtrsim 10$ Gyr) population.
Our observations were not optimized for
determining precise RRL properties, but we still have enough epochs
(24 spaced over $\sim 5$ [7] days in S1 [S2]) 
to search for and identify 
candidates based on their photometric variability.

\begin{figure*}
\includegraphics[width=\textwidth,keepaspectratio=true]{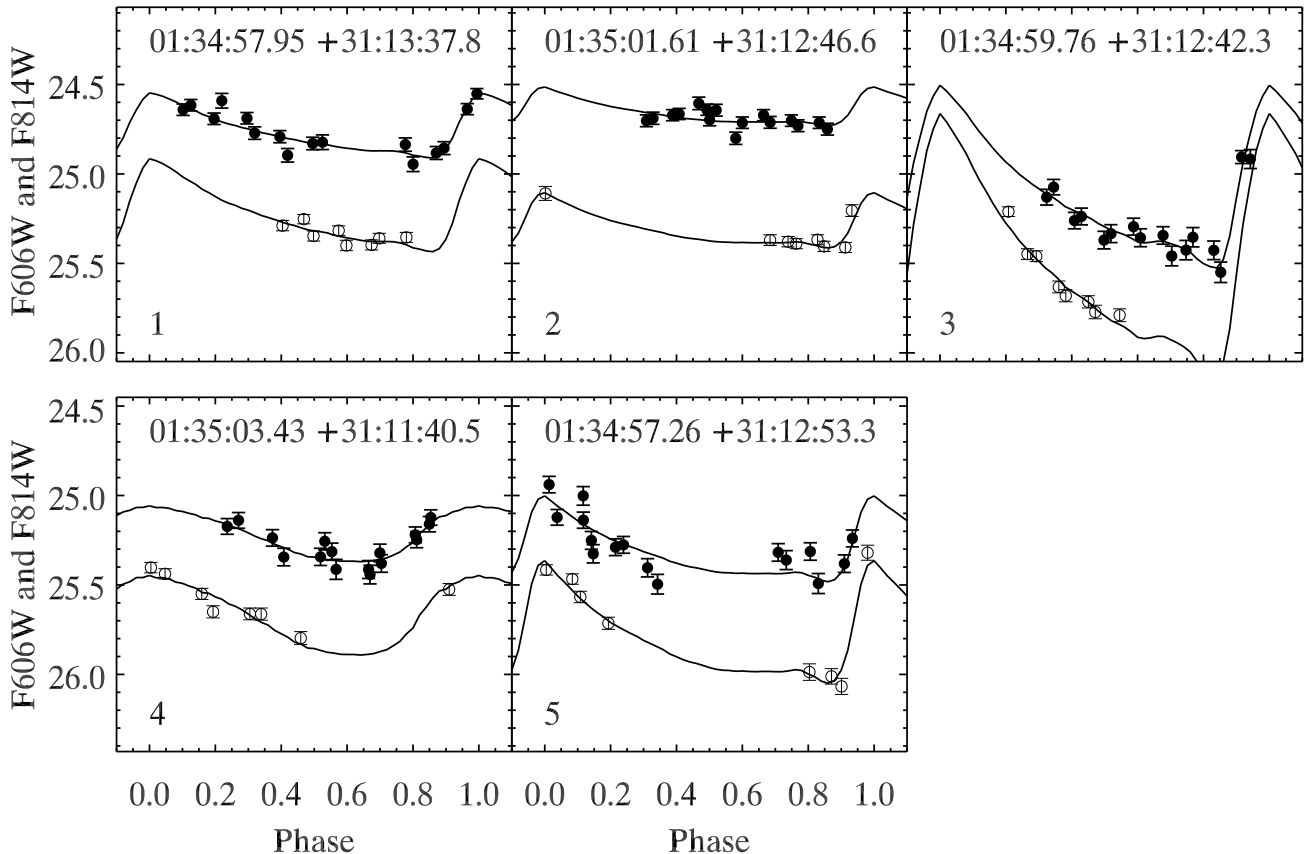}
\caption{Example template fits to the light curves of 
the 5 RRL candidates in S1.
Open points are the $F606W$-band and solid points are the $F814W$-band.
The J2000.0 coordinates are given at the top of each panel.
}
\label{fig:vars}
\end{figure*}

To that end, we used the 
\citet{Stetson96} variability index, $L$, which
measures correlated magnitude residuals 
in pairs of observations and is robust to outlier data points.
In S1, there were 5 stars 
with $L$ values $\gtrsim 3 \sigma$ above the mean 
at $F606W \sim 25 - 26$
and lying near the instability strip at 
($F606W-F814W) \sim 0.25 - 0.5$ \citep{Brown04,Bernard09}.
In S2, there were no such stars.
Fig.\ \ref{fig:vars} shows the phased light curves
and coordinates of the S1 RRL candidates after fitting 
the templates of \citet{Layden98}
and \citet{Layden00}.
The fact that we find more candidates in S1 than in S2
suggests S1 had a higher star formation rate (SFR) or a lower metallicity
at ages $\gtrsim 10$ Gyr, but given the small samples
involved and sparse phase coverage, it is difficult 
to make any firm conclusion on this matter.

\subsection{Colour Function}

\begin{figure*}
\includegraphics[width=\textwidth,keepaspectratio=true]{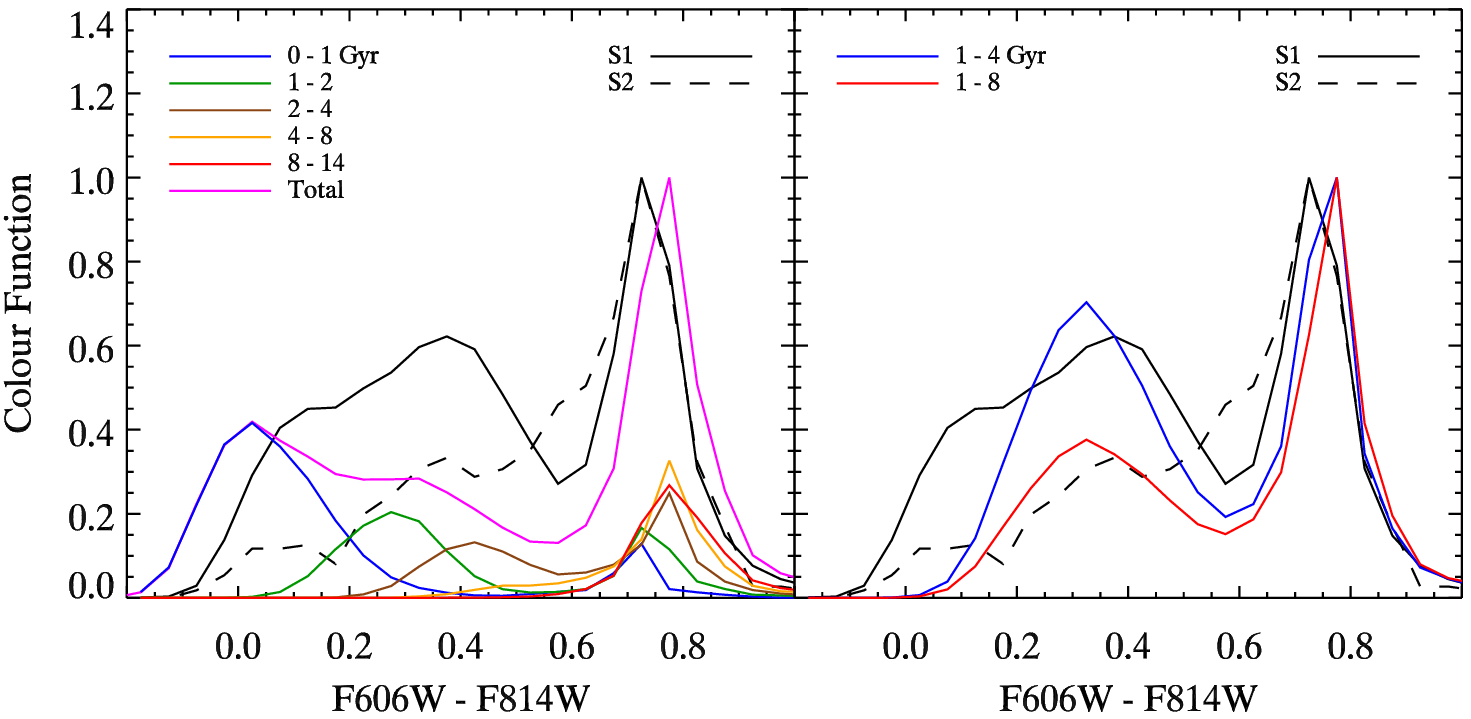}
\caption{
CF in S1 (solid black line) and S2 (dashed black line) 
summed over the
magnitude range $21 < F814W < 27$. 
The coloured lines show model CFs for different
age ranges assuming a constant SFR and [M/H] $= -0.5$ dex
and no metallicity spread.
(Left) The total model CF of all ages is represented by the
magenta line.
The data and total model CFs are normalised to
their red peaks.
(Right) The model CFs for ages 1 -- 4 Gyr and 1 -- 8 Gyr
are normalised to their red peaks.
The position of the blue peak 
and its height relative to the red peak
indicate a dominant population
with ages $\sim 1 - 4$ Gyr in S1 and an older
population in S2.
}
\label{fig:cf}
\end{figure*}

\citet{Gallagher96} and \citet{Gallart05} have shown how 
the colour function (CF) is a powerful probe of the SFH
because the main sequence turn-off (MSTO) 
and SGB temperature decreases with age.
The left panel of Fig. \ref{fig:cf} compares the CF summed over the
magnitude range $21 < F814W < 27$ 
in S1 (solid black line) and S2 (dashed black line).  
The magenta line shows a model generated with IAC-star
(see \S \ref{sec:sfhmeth} for further details) for a constant
metallicity [M/H] $= -0.5$ dex with no metallicity spread 
and constant SFR
from 0 -- 14 Gyr after convolution with the 
photometric errors and completeness.
The other lines show the contributions of several
different age ranges to the total model.

Concentrating on the left panel in Fig. \ref{fig:cf}, 
the CF for most age-groups displays two peaks: a blue peak
at $\vmi \sim 0 - 0.4$ that is associated with the 
MSTO and SGB of young and intermediate ages
and a red peak at $\vmi \sim 0.75$ that is dominated by the RC and RGB.
The data and total model CFs are normalised to
the heights of their red peaks.
The relative heights of the red and blue peaks and the position
of the blue peak vary with age and metallicity.
Ages $< 1$ Gyr contribute mostly to the blue
peak while ages $> 8$ Gyr contribute entirely to the red peak.
At intermediate ages, the blue peak shifts to redder colours
and the blue/red peak height ratio decreases.

The blue envelope in the total model is too blue and the
red peak is too red by $\sim 0.05$ mag,
perhaps indicating (1) that the metallicity should be higher
and lower for, respectively, the youngest and oldest ages, 
(2) that the bolometric corrections 
have a small zero-point offset at low or high temperatures,
or (3) that the reddening should be higher for blue
stars and lower for red stars.
The value [M/H] $= -0.5$ dex provides a reasonable
match to the entire CF and is useful for
illustrating the dominance of intermediate ages.
These considerations do not significantly
change if the metallicity varies by $\sim 0.3$ dex, 
which would mainly act to shift
the CF left or right rather than change the relative peak heights.  
Because a constant metallicity with zero spread may not apply to
the system's entire chemical enrichment history, 
we also tried a piecewise linear age-metallicity relation
increasing from --1.2 dex at 14 Gyr, to --0.7 dex at 8 Gyr, and
to --0.5 dex at the present-day with a uniform metallicity
spread of $\pm 0.1$ dex.  The resulting model CF was
slightly more similar to the original in that the
position of the red peak more closely matched the data.

As the left panel of Fig.\ \ref{fig:cf} shows, 
a constant SFR provides a relatively poor fit to the
CF of both S1 and S2.
Any arbitrary SFH can be approximated by scaling the
age-group curves accordingly and then renormalising
the sum to the red peak's new height.
For example, an exponentially increasing 
SFR from 14 Gyr ago to the present
would preferentially raise younger age-group curves 
by larger amounts and make the blue envelope of the CF too high.
In S1, the position of the blue peak and its height
relative to the red peak are most closely matched by the
1 -- 4 Gyr age range.
This is more clearly seen in the right panel of 
Fig.\ \ref{fig:cf}, which shows the CFs
for ages 1 -- 4 Gyr and 1 -- 8 Gyr after normalising
them to their red peaks.
Crucially, this means that ages 1 -- 4 Gyr 
can account for the majority of the SGB, RC, and RGB stars
in S1 and that there is
little room for stars older than 8 Gyr, which
would push up the red peak without changing
the height of the blue peak.
The lack of a strong blue peak in S2 again suggests 
an older population than in S1.

\subsection{Vertical Clump Morphology}
\label{sec:vclump}

\begin{figure*}
\includegraphics[width=\textwidth,keepaspectratio=true]{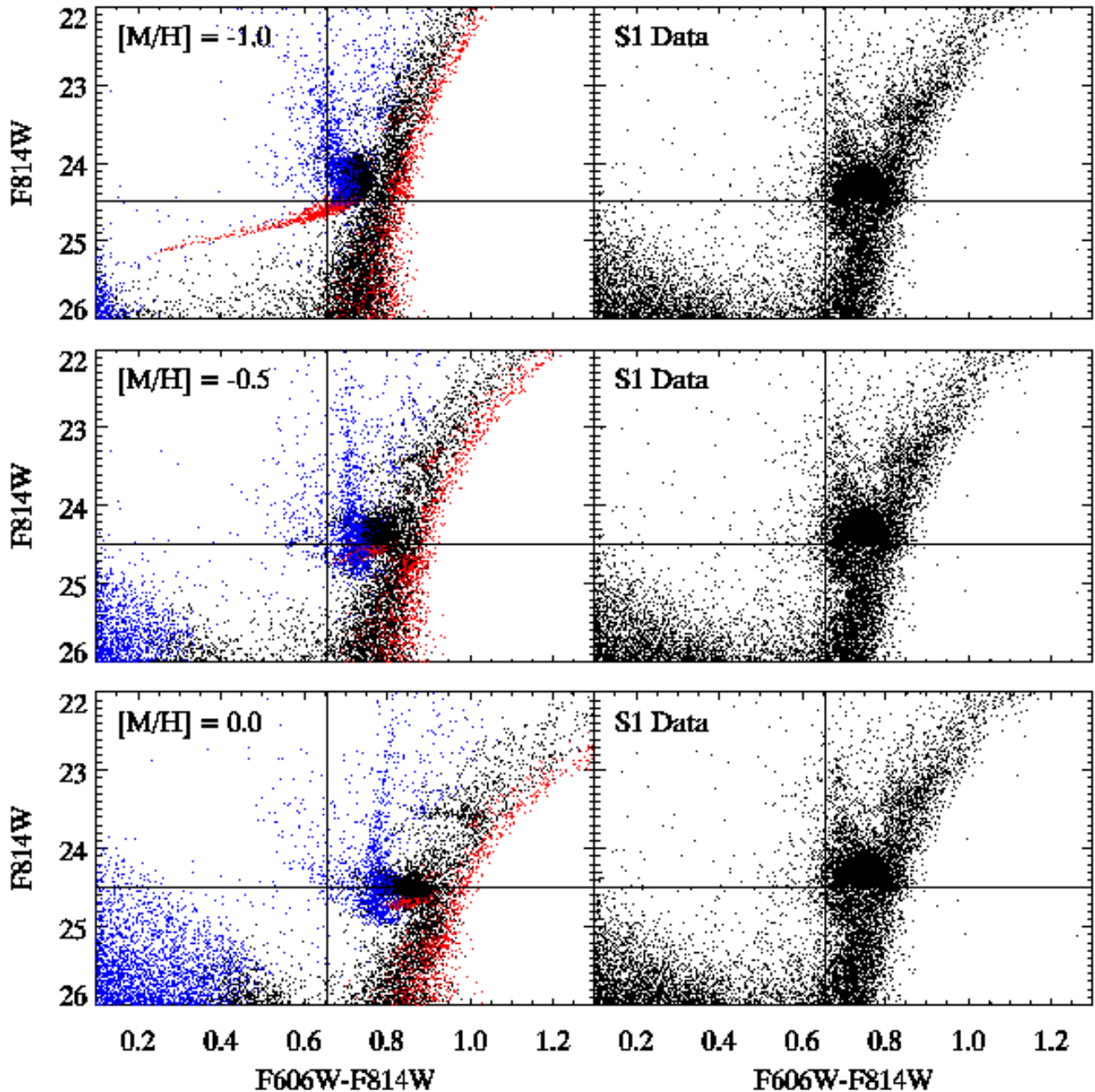}
\caption{Using the morphology of the vertical clump to
constrain metallicity.  The right-hand column shows the data
for S1 and the left-hand column shows, from top to bottom, synthetic CMDs for
[M/H] = --1.0, --0.5, and 0.0 dex and constant SFR.
The points in the synthetic CMDs are divided into the age ranges 
$8 - 14$ Gyr (red), $1.2 - 8$ Gyr (black), and $<1.2$ Gyr (blue).
The solid lines are meant to guide the reader's eye.
The morphology of the vertical clump and its extension 
below the RC are best matched by [M/H]$ \sim -0.5$ dex.
}
\label{fig:rheb}
\end{figure*}

Finally, we examine how a particular feature of the S1 CMD can constrain
the metallicity of young stars in S1.
At ages of $\sim 0.4 - 1.2$ Gyr, 
stars with masses in the range $\sim 2 - 3\ M_{\sun}$ form
a vertical clump that extends from and partially overlaps 
with the left-hand side of the RC.
The slope of this vertical clump depends on metallicity 
while its extension below the RC appears only for 
$[M/H] \gtrsim -0.7$ dex \citep{Gallart05}.
Fig.\ \ref{fig:rheb} shows how the morphology of the
vertical clump constrains the metallicity at these ages.
In the left-hand column are three synthetic CMDs
generated with IAC-star 
(see \S \ref{sec:sfhmeth} for further details)
assuming a constant SFR from 0 -- 14 Gyr.
From top to bottom, the panels in the left-hand column assume
a constant metallicity of [M/H]$ = $--1.0, --0.5, and 0.0
dex with no metallicity spread.  
The right-hand column shows the S1 CMD.
The solid lines in each panel are meant to guide the reader's eye.
The synthetic stars are colour-coded by three age ranges, 
0 - 1.2 Gyr (blue), 1.2 - 8 Gyr (black), and $>$ 8 Gyr (red).
This figure demonstrates that the slope, length, and position of the
vertical clump are best matched by [M/H] $\sim -0.5$ dex, 
as is the position of the blue plume in the lower-left
corner of each panel.
The match is not perfect, but this is not surprising 
since the stellar tracks or bolometric corrections 
may have zero-point offsets in colour and magnitude
and, for this test, the metallicity was constant with no spread.
It is also worth noting that the slope of the RGB and
morphology of the RC are best matched by the same
metallicity, especially if we exclude stars older than 8 Gyr.

In summary, various features of the S1 CMD point to
a dominant population with ages $\sim 1 - 4$ Gyr, 
and, at younger ages, a metallicity $> -0.7$.
The fainter SGB, but similar RC magnitude and
RGB colour in S2 indicate an older and more metal-poor
population.

\section{Star Formation History}
\label{sec:sfh}
\subsection{Method}
\label{sec:sfhmeth}

To obtain a more detailed look at the 
mix of ages and metallicities in these fields,
we used the method of synthetic CMD fitting.
This method involved fitting the observed CMD with 
a linear combination of individual basis populations
each representing a range of ages and metallicities.
The coefficients of the linear combination gave the SFRs
at their respective ages and metallicities.
Various implementations of this technique have been used
extensively over the years 
\citep[e.g.,][]{Tosi91,Bertelli92,Tolstoy96a,Aparicio97,
Gallart99,Hernandez00,Harris01,Holtzman99,Dolphin02,Cole07,Aparicio09}
and in this paper we
followed closely that of \citetalias{Barker07b}.

The basis population CMDs were generated with IAC-star \citep{Aparicio04} 
and then convolved with the photometric errors and
completeness rates derived from the artificial star tests.
The entire set of basis CMDs spanned ages from
25.1 Myr to 14.1 Gyr and metallicities from --1.7 to +0.1 dex.
The age range was broken up into 13 bins whose widths increased
with age because the photometric spacing between stellar isochrones decreases
as they get older.
The metallicity range was divided into 6 uniform bins each 0.3 dex wide.
We adopted the BaSTI theoretical stellar evolutionary tracks 
\citep[][called Teramo in IAC-star]{Pietrinferni04}
with mass loss parameter $\eta = 0.4$ 
and ACS bolometric corrections.
We note that using the \citet{Girardi00}
tracks gave similar results
with a tendency for a mean overall metallicity $\sim 0.2$ dex 
lower than the BaSTI tracks.
We take this is an indication of the systematic error
on our metallicities.

For the IMF, we adopted a Salpeter law from 0.1 to 120 $M_{\sun}$.
Other IMFs that turn over at low masses would mainly
act to change the overall normalization of the SFH rather
than its shape.
The other input parameters were the binary fraction
and minimum binary mass ratio, both of which we set to 0.5.
\citet{Gallart99} summarized observational evidence 
supporting similar values.
Tests whereby we fit mock populations with known parameters 
did not reveal any systematic effects that might
arise from an incorrect binary fraction.
The fitting region in the CMDs spanned the colour
range $-0.5 < (F606W-F814W) < 2.0$
and the magnitude rage $21 < F814W < 27$
where both fields are $\gtrsim 60\%$ complete.
Including fainter magnitudes resulted in poorer fits, 
but broadly similar SFHs.
Within the fitting region, the data and model CMDs
were divided into square boxes 0.1 mag on a side.

We tried masking the RC and upper RGB ($F814W < 25$) from the fits 
and the solutions were again similar to the originals
using both the BaSTI and Girardi tracks.
We also found that the differences between the 
BaSTI and Girardi solutions were smaller
when including the RC and upper RGB in the fits.
We think that this is because, by using more of the 
CMD regions available, we effectively
average over their systematic errors 
and are not as vulnerable to the errors in any one
particular region.
It is possible, however, that a different approach
would be necessary for deeper data or 
other stellar systems populating 
different regions of the age-metallicity plane.
In all the tests we conducted, the fraction of stars formed
over 4.5 Gyr ago in both fields differed from the original BaSTI solution
by $\lesssim 10\%$ and the mean age differed by a few tenths of a Gyr.
To be conservative, we take 1 Gyr as the systematic uncertainty
on the mean age.

We used the StarFISH software package \citep{Harris01}
to find the best-fitting model 
according to a maximum likelihood statistic appropriate for
Poisson-distributed data \citep{Dolphin02}.
StarFISH uses a downhill simplex algorithm
to search through parameter space and locate the best fit.
The stock implementation of this algorithm employs a 
random search through parameter space to locate a 
reasonable starting position for the simplex.
Instead of using this random search, 
we initialized the simplex to 
the best-fitting solution found by the genetic
algorithm PIKAIA \citep{Charbonneau95}.
For consistency, we applied this hybrid approach to
both fields although it made no difference to the
solutions for S1.
When applied to S2, the hybrid approach 
was less prone to getting stuck in local minima, but
the solutions were consistent with those of the stock
implementation to within the errors.

StarFISH computed the error bars in a 3-stage process 
designed to measure the independent as well as correlated
errors between coefficients \citep{Harris01,Harris04}.
First, individual coefficients were varied while holding all others
fixed at their best-fit values.
Second, adjacent pairs of coefficients were varied 
holding the rest fixed at their best-fit values, and, 
third, all coefficients were varied simultaneously.
At each stage, the program iteratively stepped away from
the best-fit and computed the new likelihood ratio
until this ratio reached its $1\sigma$ limit.
Throughout the whole process, the program 
updated and stored the 
maximum variation of every coefficient resulting
in the $1\sigma$ limit.
The final maximum variations were taken as the 
$1\sigma$ confidence intervals
for the coefficients.

As a zero-order correction to any possible errors
in the stellar tracks or bolometric corrections, we also
solved for distance and extinction.
The distance modulus was varied from 24.40 -- 24.90 in steps of 0.1
and the $F606W$ extinction was varied from 0.00 -- 0.30 in steps of 0.05.
All acceptable solutions on this grid were averaged together and
their dispersion was added in quadrature with the error bars
of the best individual solution.
With this procedure, the final error bars include 
random and correlated errors of the SFH amplitudes as well
as zero-point uncertainty in the stellar tracks
and bolometric corrections.
The error bars do not necessarily reflect the variation
of the true SFR within an age bin, but rather
the $1\sigma$ confidence interval on the SFR
averaged over an age bin.

\subsection{Results}
\label{sec:sfhres}

\begin{figure*}
\includegraphics[width=\textwidth,height=22cm,keepaspectratio=true]{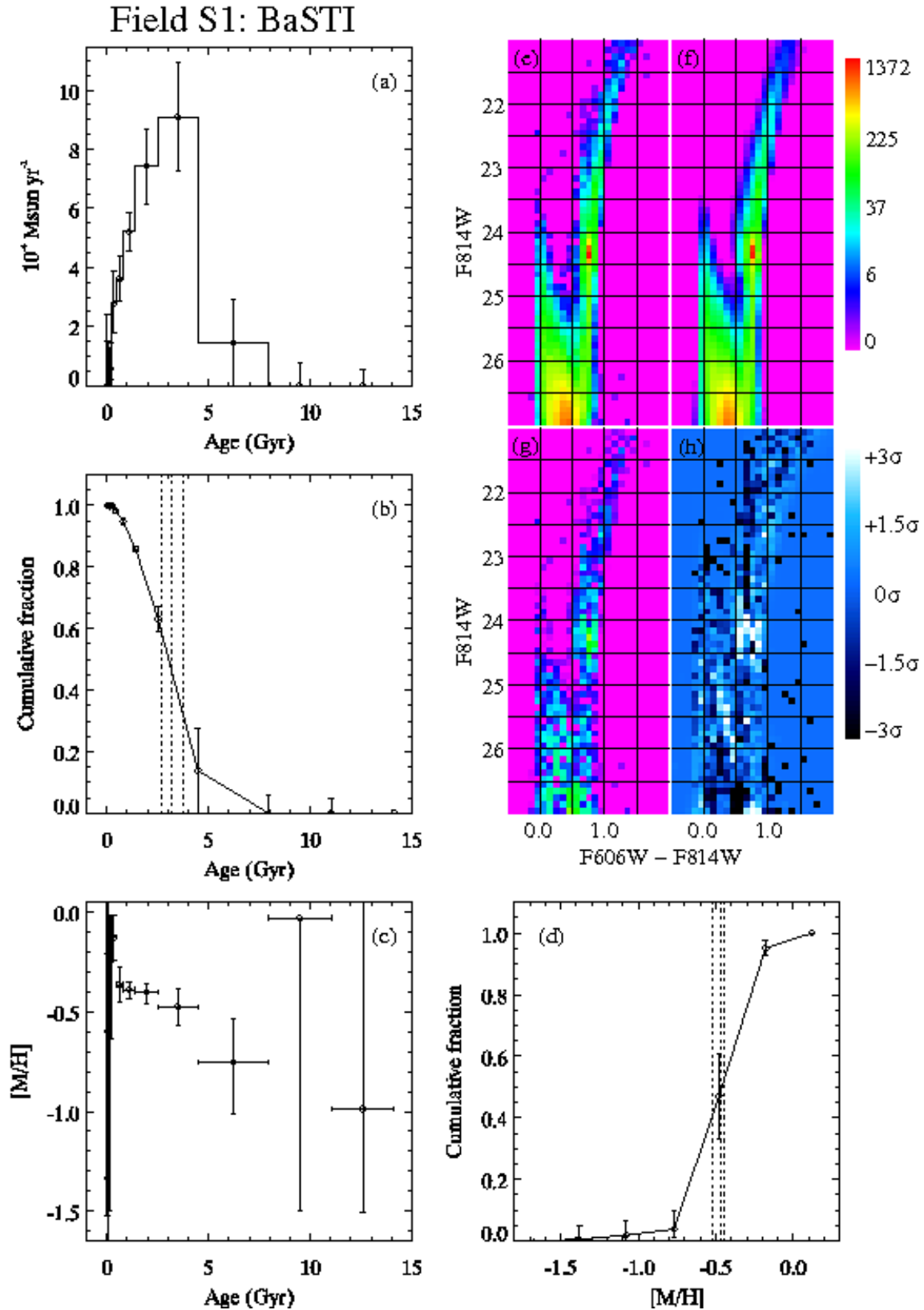}
\caption{The SFH of field S1.
The panels show (a) the SFH, 
(b) the cumulative age distribution, 
(c) the age-metallicity relation, 
(d) the cumulative metallicity distribution, 
(e) the data CMD, (f) the model CMD, 
(g) the absolute magnitude of the residuals,
and (h) the significance of the residuals.  
Panels (e) -- (g) use the same 
logarithmic intensity scale.
The intensity scale in (h) goes from $-3\sigma$ (black)
to $+3\sigma$ (white) where negative values indicate
the model is too low.
The vertical dotted lines in (b) and (d) are mean values and their
$1\sigma$ confidence intervals.
}
\label{fig:S1_teramo}
\end{figure*}

\begin{figure*}
\includegraphics[width=\textwidth,height=22cm,keepaspectratio=true]{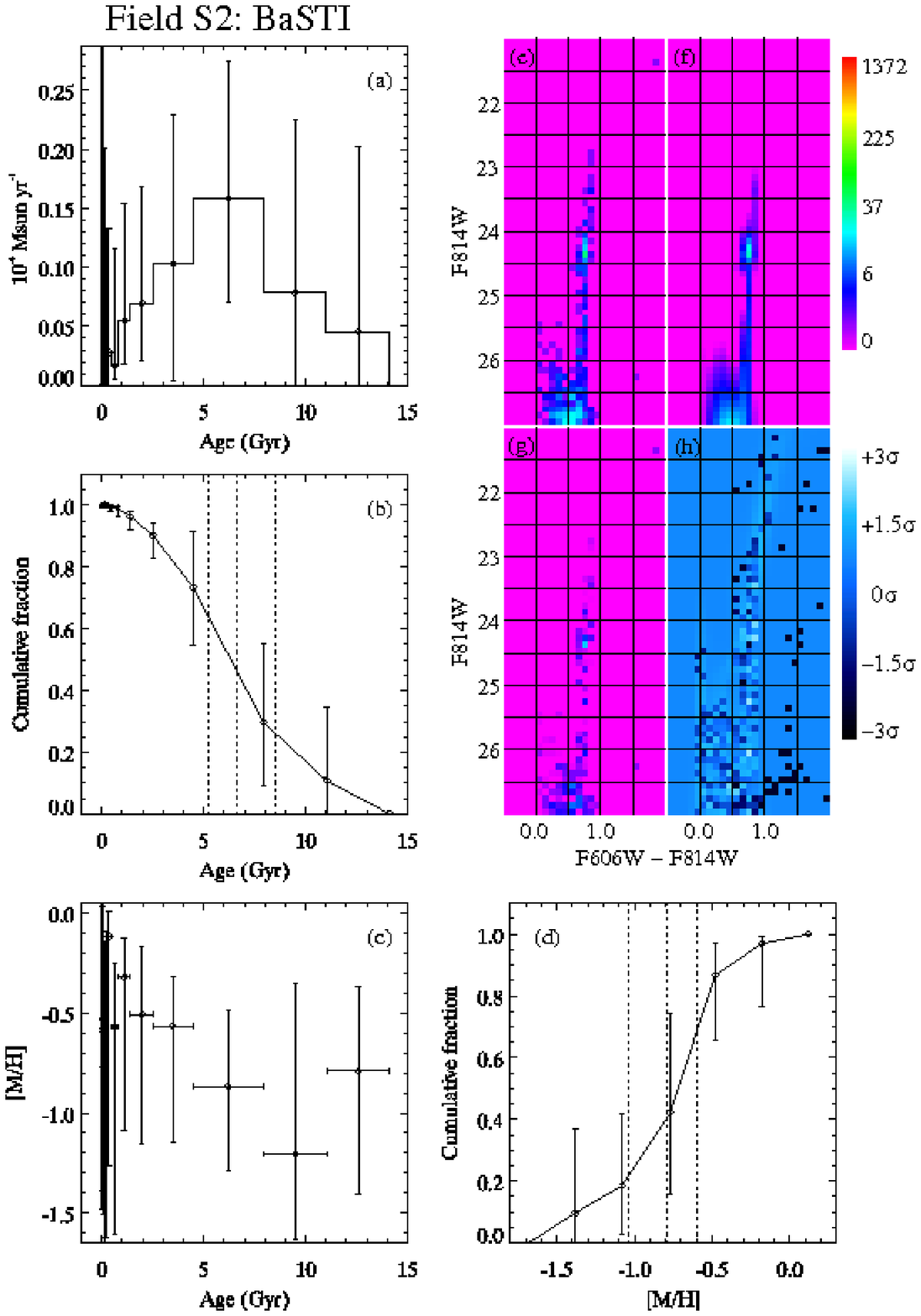}
\caption{
The SFH of field S2.  The panels show the same information as 
Fig. \ref{fig:S1_teramo}.
Note that the y-axis range in panel (a) is reduced
compared to Fig.\ \ref{fig:S1_teramo}.
The $1\sigma$ upper limits on the SFRs in the first, second,
and third youngest age bins in panel (a) 
extend off the top of the graph.  Their values are 
$\rm 0.9,\ 0.5,\ and\ 0.3\ \times 10^{-4}\ M_{\sun}\ yr^{-1}$, respectively.
}
\label{fig:S2_teramo}
\end{figure*}

Figures
$\ref{fig:S1_teramo} - \ref{fig:S2_teramo}$
present the results of the CMD fitting analysis.  
The recovered SFHs are in line with our qualitative 
expectations from \S \ref{sec:cmds}.
In field S1, the mean age is $3.2$ Gyr and
most of the star formation took
place $\sim 2 - 4$ Gyr ago, coinciding with a general peak in the SFR.
The decline in the SFR over the past 1 Gyr is a robust result because
a constant SFR would produce too many stars on the
blue plume at $\vmi \sim 0.0$ as found from the CF analysis.
The age of the SFR peak is similar to that 
measured by \citetalias{Barker07b}, but
we must be careful not to over-interpret 
this feature in the solutions.
Any CMD fitting analysis is limited in its 
ability to measure the burstiness
of the underlying SFH.  
Short bursts with timescales smaller than the
age bins will be heavily smoothed.
Longer bursts occurring over several age bins
will be more reliably recovered, but the recovered 
peaks may drift by one 
age bin due to the finite model binning 
and errors in the model input parameters.
We found that using smaller age and metallicity bins or
shifting the bins by a fractional amount 
did not change the general picture of a period of enhanced
SFR $\sim 2 - 4$ Gyr ago
and had only a small effect on the mean age, 
at the level of $\sim 0.5$ Gyr.

In field S1, the mean SFR over all ages is $2.6 \times 10^{-4}\ \msunyr$
and the mean SFR at ages $<$ 8 Gyr is $4.5 \times 10^{-4}\ \msunyr$.
The total stellar mass formed is $3.6 \times 10^{6}\ \msun$
resulting in a projected stellar surface density of 
$\sim 5\ \msunpc$.
Correcting for an inclination of $56\degr$ yields 
a stellar surface density of $\sim 2.7\ \msunpc$
in the disc plane.
Repeating the fit with an IMF that turns over at $0.5\ \msun$
\citep{Kroupa02} gave a stellar surface density
$\sim 25\%$ lower, but the relative age distribution was unchanged.
The best-fitting distance modulus is
$24.72 \pm 0.09$, in good agreement with the \citet{Galleti04}
value.  The best-fitting extinction is $A_{F606W} = 0.05 \pm 0.03$,
lower than the \citet{Schlegel98} value of 0.12.
This difference probably reflects a zero-point
offset in the stellar evolutionary tracks or
bolometric corrections as was hinted at in the
analysis of the vertical clump feature in
\S \ref{sec:cmds} and in Fig.\ \ref{fig:rheb}
or it could reflect small scale reddening
variations not captured by the \citet{Schlegel98} maps.


The cumulative age distribution in panel (b) is actually more stable
against correlations between age bins than the differential SFH
in panel (a).
Roughly $85\%$ of stars in S1 formed by 1.5 Gyr ago and
$\sim 65\%$ formed by 2.5 Gyr ago.
Only $14\%$ of stars have ages $> 4.5$ Gyr, 
but this value could be as low as $0\%$ or as high as $28\%$ 
and still provide a fit consistent with the original at the $1\sigma$ level.  

There is a trend of decreasing metallicity 
with age between 0.35 and 6.2 Gyr in S1
as expected in normal 
chemical evolution scenarios.
However, the trend is weak and consistent with
no evolution over most of the history,
and it partially relies on the RC/RGB
morphology.
After repeating the fit with the RC and upper RGB masked, 
the new S1 age-metallicity relation did not 
continuously decrease from 2.0 to 6.2 Gyr,
but it was consistent with the original
at the $1\sigma$ level or differed by $< 0.2$ dex in all age bins.
Because of the weak metallicity evolution, the
metallicity distribution is narrow, with $\sim 90\%$
of stars having [M/H] between --0.8 and --0.2 dex.
The mean metallicity of all stars ever formed
is [M/H] $= -0.47^{+0.03}_{-0.05}$ dex.  These error bars
represent only the formal random error, but
the systematic error is $\sim 0.2$ dex.
The lack of stars with [M/H] $< -0.8$ dex may be
the result of pre-enrichment of the gas out of which
most stars formed.
The metallicity at 0.6 -- 1.1 Gyr is $\sim -0.4$ dex, 
in reasonable agreement with our analysis 
of the vertical clump morphology in \S \ref{sec:vclump}.
Because there is such a small signal in the oldest 2 bins, 
their metallicities are completely unconstrained.
Likewise, there are not enough stars at 
ages $< 0.35$ Gyr to populate 
the metallicity-sensitive core helium burning
(blue loop) phase, and, consequently, the 
age-metallicity relation is also unconstrained
at these ages.

The metallicity in S1 drops 
from $-0.13 \pm 0.11$ dex at 0.35 Gyr
to $-0.36 \pm 0.10$ dex at 0.62 Gyr.
The latter value may be a better indicator of the
present-day metallicity because it is constrained by the
metallicity-sensitive
position and morphology of the vertical clump 
whereas the former value comes primarily from 
the colour of the upper MS, which varies little with metallicity.
Extrapolation of the metallicity gradient in M33 measured by \citet{U09}
from A- and B-type supergiants at $R_{dp} \lesssim 7$ kpc
predicts a lower present-day value in S1, [M/H] $= -0.65 \pm 0.13$ dex.
The HII region and planetary nebula gradients computed 
by \citet{Bresolin10} predict $[O/H] = -0.54 \pm 0.10$ dex
and $[O/H] = -0.34 \pm 0.20$ dex, respectively, when extrapolating
beyond their outermost measurements at $R_{dp} \sim 8$ kpc and adopting
the solar oxygen abundance of \citet{Asplund09}.
Using the solar oxygen abundance of \citet{Grevesse98}
would shift their measurements down by $\sim 0.1$ dex.
The discrepancy between our measurement and theirs
could mean that the metallicity
gradient flattens out 
beyond the outermost measured 
supergiants, HII regions, and planetary nebulae, or that
M33's inner disc has accreted metal-poor gas 
in the last 0.35 Gyr. 

Compared to field S1, S2 is older and more metal-poor, 
but the small number of stars in this field limit the
statistical significance of this result.
The interquartile age range in S2 is $\sim 4 - 9$ Gyr, 
$\sim 75\%$ of stars have ages $>$ 4.5 Gyr, 
and $\sim 90\%$ have ages $>$ 2.5 Gyr.
Most stars have metallicities between --1.0 and --0.5 dex.
The mean SFR over all ages is $8.9 \times 10^{-6}\ \msunyr$.
The total stellar mass formed is $\sim 30$ times
less than in S1, or $1.2 \times 10^{5}\ \msun$,
resulting in projected and deprojected stellar surface densities of 
$\sim 0.18\ \msunpc$ and $\sim 0.10\ \msunpc$, respectively, 
assuming the same inclination as in S1.
The best-fitting distance modulus is
$24.69 \pm 0.09$ and the best-fitting
extinction is $A_{F606W} = 0.04 \pm 0.05$, in good
agreement with those found for field S1.

In \S \ref{sec:cmds}, we identified 5 RRL candidates
in S1 and none in S2.
Are the derived SFHs for S1 and S2 consistent with
this finding?
In field S1, the $1\sigma$ upper limit 
on the SFR in the oldest age bin at 12.6 Gyr 
yields $\sim 1.7 \times 10^{5}\ M_{\sun}$ of integrated star formation.
In S2, this value is $\sim 5.0 \times 10^{4}\ M_{\sun}$.
From the synthetic CMDs in our library, we estimate 
that a $10,000\ M_{\sun}$ population 
with age $\gtrsim 8$ Gyr and $[M/H] \lesssim -0.4$
contains $\sim 4$ HB stars 
regardless of HB morphology and with little
dependence on metallicity.
This yields a $1\sigma$ upper limit of $\sim 70$ HB stars in S1 and
$\sim 20$ HB stars in S2.
Given that we expect the RRL
to be a subset of the total number of HB stars, 
the number of RRL candidates we found in S1 and S2
is well within the $1\sigma$ uncertainty
in the SFR of the oldest age bin.

\section{Discussion}
\label{sec:disc}

\begin{figure}
\includegraphics[width=80mm,keepaspectratio=true]{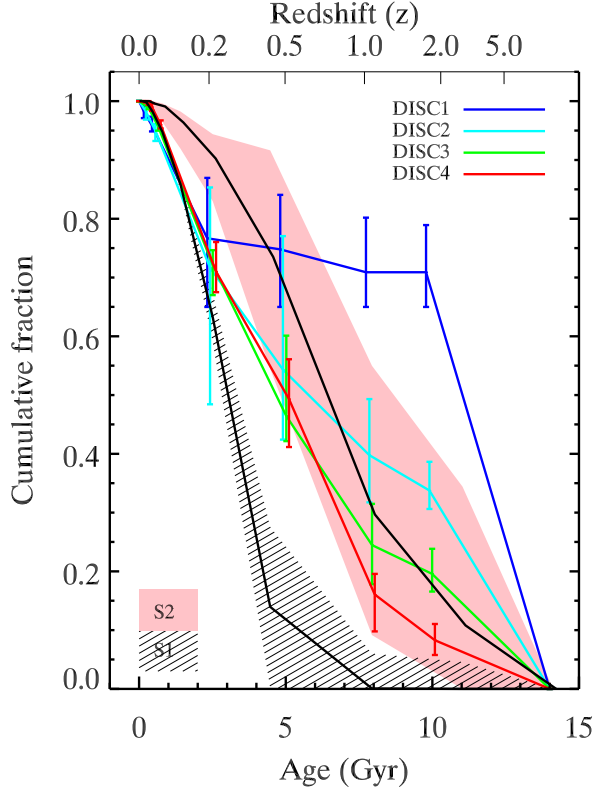}
\caption{
Cumulative age distributions for fields S1 
and S2 (black lines).
Their $1\sigma$
confidence intervals are displayed as shaded regions
assuming a constant SFR within each age bin.
From top to bottom, the four solid lines with error bars are the
results of \citetalias{Williams09b} covering $R_{dp} \sim 1 - 6$ kpc.
Field S1 is younger than their outermost field (DISC4).
Field S2 is older than S1 and this difference is not just in
the recent SFH, but likely stretches back over their entire histories.
}
\label{fig:cads}
\end{figure}

The CMD fitting analysis indicates that 
almost the entire stellar mass in S1 formed in the last 8 Gyr
and that field S2 is older and more metal-poor.
Fig.\ \ref{fig:cads} 
compares the cumulative 
age distributions of both fields.
The shaded regions represent the $1\sigma$ confidence intervals assuming a
constant SFR within each age bin.
From this comparison, it is likely that S2 has always
been older than S1, but 
given the depth of our photometry, we cannot say
precisely when star formation started in each field.
The $1\sigma$ errors allow for nearly the same fraction of 
stars $> 8$ Gyr old in both S1 and S2, so it is feasible
that star formation in S1 started at the same time as
or even before it did in S2.
However, the fraction of ancient stars is unlikely to be significantly
larger in S1 than in S2.

Fig.\ \ref{fig:cads} also shows the results of 
\citetalias{Williams09b} for their 4 fields at
$R_{dp} \sim 1 - 6$ kpc.
The cumulative star formation in S1 is less than 
in the \citetalias{Williams09b} outermost field (DISC4)
at nearly all ages, demonstrating that the
inside-out growth of M33's inner disc 
extends all the way to the disc edge.
In addition, the difference in SFH between S1 and S2
adds further support to the idea that M33's
age gradient reverses at large radii
near the break radius.

\subsection{Explanations for M33's Age Gradient}

There are several possible explanations 
for the behavior of M33's mean age profile.
The one favoured by \citetalias{Williams09b} involved the secular redistribution
of stars seen in recent simulations of
isolated disc formation and evolution \citep{Roskar08b}.
In these simulations, inside-out disc growth leads to a negative age
gradient within the break radius, but radial mixing of stars
due to interactions with transient spiral density waves 
causes a positive age gradient beyond the break radius.

If this scenario is true, then most of the stars in S2
formed at smaller radii and the ages of the youngest stars in S2
could constrain the stellar migration timescale.
Inspection of the S2 CMD and SFH
shows that the youngest stars are likely
to have ages $< 1$ Gyr.
There are a few stars that lie near the 
200 Myr isochrone in the right panel of Fig. \ref{fig:cmdiso}, 
but according to the Besancon model for the MW \citep{Robin03}, the
brightest of these (at $F814W \sim 22$) may be foreground stars.
A migration timescale of $\sim 0.2 - 1$ Gyr 
appears consistent with the simulation of \citet{Roskar08b}, 
in which the disc break appears quickly. 
However, the youngest stars, which represent a small fraction of the
total in S2, may have formed in-situ.
If so, then they formed at sub-threshold gas densities.
Typical empirical and theoretical estimates of the
star formation threshold lie in the range $\sim 3 - 10\ \msunpc$
\citep{Skillman87,Taylor94,Ferguson98,Martin01,Elmegreen94,Schaye04},
but the HI density in S2 is only $0.1\ \msunpc$
\citep[][D.\ Thilker, private communication]{Thilker02}.
In comparison, field S1 is more gas-rich
with an HI density of $4.1\ \msunpc$
\citep[][D.\ Thilker, private communication]{Thilker02}, so 
the youngest stars in S1 could have 
formed in-situ without violating previous estimates of the
SF threshold.

The radial mixing in the above scenario
can cause the measured SFH in any 
one region of a galaxy to differ from the true one,
especially in the outermost disc regions.  
\citet{Roskar08a} found a nearly constant measured 
SFH near the break radius of their simulated galaxy whereas the true 
SFH was significantly skewed toward younger ages.  
Our measured SFH in S1 strongly rules out a constant 
SFH and is actually more consistent with their true SFH.  
In addition, field S1 is unlikely to
be significantly contaminated by stars that migrated from 
the inner disc because of the small 
fraction of stars older than 4.5 Gyr 
($\sim 14\%$) relative to the inner disc 
where such stars are common \citepalias[$\gtrsim 50\%$; ][]{Williams09b}.

Aside from stellar radial mixing, 
there are other possible explanations for
the behaviour of M33's mean age profile.
\citet{SanchezBlazquez09} studied the formation of
an early type spiral galaxy in a high resolution
cosmological simulation.
This galaxy exhibited a break in the exponential radial 
SB profile analagous to that in M33, 
but the stellar mass density profile
exhibited no such break
because of a net radial migration of stars
towards the outskirts.
This galaxy also had 
an upturn in the age gradient
at the break radius, but \citet{SanchezBlazquez09} argued that this
was caused not by the radial migration, but by a warp in the gas disc
and a concurrent drop in gas volume density and SFR.
It is well known that the gas disc of M33 begins to
warp near the edge of the optical disc
\citep{Rogstad76,Corbelli97}, lending support to this
alternative scenario.
However, contrary to the simulations of \citet{SanchezBlazquez09}
and \citet{MartinezSerrano09}, we find that M33 has a 
sharp drop in the stellar mass density near the break radius.

A change in the SF efficiency 
may also explain the upturn in M33's age profile.
Such a change might be expected to occur when
the gas density drops below the SF threshold,
which seems to occur near M33's break radius today.
A change in SF efficiency would change the relationship
between gas and SFR, behaviour that has been observed
in the outskirts of spiral galaxies where the gas density is low
\citep{Bigiel08,Leroy08}.
Moreover, \citet{Leroy08} observed a correlation
of the star formation efficiency with stellar mass surface density.
Thus, we would naturally expect
to see a difference in SF
efficiency and SFH between regions inside
the break radius and regions outside it.

The older mean age 
beyond M33's break radius could also be due to 
a transition from its thin disc to its halo or thick disc.
In this case, the mean age
of $\sim 6 - 8$ Gyr found in this work and in B07 
for the region beyond the break 
would imply that the halo or thick disc is younger 
than it is in the MW, where these components are predominantly
$> 10$ Gyr old \citep{Gilmore95,Krauss03}.  
The mean metallicity of $-0.7$ to $-0.9$ 
outside the break radius found in this study and in \citetalias{Barker07b}
is higher than estimates for M33's halo field stars
and globular clusters, 
which lie in the range $-1.5$ to $-1.3$ \citep{Sarajedini00,McConnachie06}.
It is not known whether M33 contains a thick disc, 
but studies of the diffuse light around galaxies
have found that thick discs may be common 
\citep[e.g.][]{Burstein79,Morrison97,Neeser02,Dalcanton02}.
The mean age and metallicity in S2 are
in the range found by \citet{Yoachim08b}
for thick discs of several late-type galaxies
using Lick indices.

\subsection{Outer Disc Age}

Finally, the results presented here suggest that the
last major epoch of SF in M33's outer disc occurred
$\sim 2 - 4$ Gyr ago, or at $z \sim 0.2 - 0.4$
for a standard WMAP $\Lambda \rm CDM$ cosmology \citep{Dunkley09}.
This provides a unique test of the disc assembly
history predicted by cosmological N-body/SPH simulations.
Most simulations until recently have focused 
on massive early-type spirals which may have had more
active merger histories than M33, but
the physical processes and timescales governing 
thin disc growth may be similar for a range of
galaxy masses \citep{Brooks09}.

In the simulation of \citet{Abadi03}, most 
stars in the thin disc formed over the 
last $\sim 8$ Gyr via the smooth, fairly constant accretion 
of cold gas originating in accreted satellites or 
dense intergalactic filaments.  The thin disc mean 
age was $\sim 7$ Gyr at $R = 1$ kpc and $\sim 3$ Gyr at 
$R = 20$ kpc (or 4 disc scale-lengths).  
Our results for S1 are in good agreement
with theirs at the same disc location in terms of 
disc scale-lengths, but S1 is younger than the
equivalent location in their disc measured in kpc.

\citet{SommerLarsen03} reported an exponentially
declining rate of gas accretion onto their two
simulated spiral galaxies.
One galaxy formed 
inside-out, but had very little age gradient with a 
mean stellar age of $\sim 7 - 8$ Gyr. 
The other galaxy formed outside-in, with a 
mean age increasing from $\sim 4$ Gyr at 2 disc 
scale-lengths to $\sim 6.5$ Gyr at 6 disc scale-lengths, 
or R = 10.5 kpc. 
The authors did not decompose these ages into
thin disc, thick disc, and halo contributions, 
so the ages are strictly upper limits for the thin disc, 
but taken at face value, they are older than 
what we find in S1.

Robertson et al. (2004) employed 
a multi-phase ISM in their cosmological simulation, 
resulting in the formation of a bulgeless galaxy 
with total mass similar to M33.
However, their galaxy was older 
than M33 and did not simply form inside-out. 
The mean stellar age increased from $\sim 7.5$ Gyr 
in the nucleus to $\sim 10$ Gyr at 10 kpc, or 3 scale-lengths.
Moreover, the stellar disc of their galaxy appeared much thicker
than those of normal late-type spirals.

\citet{Governato09} examined the formation of a 
bright, disc-dominated galaxy in a high resolution
cosmological SPH simulation.
The last major merger occurred about 6 Gyr ago, 
after which time an extended
thin disc formed through the accretion of cold gas
from the cosmic web and cooled halo gas.
This formation timescale is qualitatively consistent
with what we find in M33's outer disc.
Stars that formed prior to the last major merger
ended up at $z = 0$ in a thick disc component with
similar mass as the thin disc, but shorter scale-length.

\section{Conclusions}
\label{sec:conc}

We have analysed two HST/ACS 
fields at 9.1 and 11.6 kpc, 
straddling the break in the SB profile on 
M33's northern major axis.
These observations offer the deepest view yet of the
stellar populations in the outskirts of M33, 
providing a valuable observational constraint on 
cosmological galaxy formation simulations.
Based on a CMD fitting analysis, we find that the
majority of stars in both fields combined formed at $z < 1$
and that the last major epoch of 
star formation at the edge of M33's disc occurred at $z \sim 0.2 - 0.4$.

The mean age in the inner field, S1, is $\sim 3 \pm 1$ Gyr
and the mean metallicity is [M/H] $\sim -0.5 \pm 0.2$ dex.
Approximately half of all stars in S1 have ages of 2.5 -- 4.5 Gyr
and only $\sim 14 \pm 14\%$ have older ages.
The SFH in S1 unambiguously reveals how the
inside-out growth of M33's inner disc \citepalias{Williams09b}
extends all the way to the disc edge.
In comparison, the outer field, S2, 
is older (mean age $\sim 7 \pm 2$ Gyr), 
more metal-poor (mean [M/H] $\sim -0.8 \pm 0.3$ dex)
and contains $\sim 30$ times less stellar mass.

These results 
provide the most compelling evidence yet that M33's
age gradient reverses at large radii
near the break radius.
As noted by \citetalias{Williams09b}, 
this behaviour is generally consistent 
with the simulations of \citet{Roskar08b}, 
in which the inner disc forms inside-out and the
region beyond the break is populated with stars that migrated from
the inner disc.
If this scenario is correct, 
the radial mixing could in principle contaminate the sample in S1 with
stars that migrated from the inner disc.
However, we argue that this effect is likely to be 
small given the small fraction of stars older than 4.5 Gyr
in S1 ($\sim 14\%$) relative to the inner disc 
where such stars are common ($\gtrsim 50\%$).

There remain alternative explanations for
the behaviour of M33's mean age profile that we cannot
completely rule out.
One explanation involves a drop in the SFR
caused by a warping of the gas disc \citep{SanchezBlazquez09}.
Another involves a change in the SF efficiency,
perhaps caused by the gas density
dropping below a critical value 
or by the sharp drop in stellar mass at the break radius
\citep{Bigiel08,Leroy08}.
A third explanation involves a transition to another 
component with a different SFH than the inner disc,
such as a halo, thick disc, or accreted substructure.
Kinematic information for individual stars in S2 as well as
a global comparison between the gas and stellar surface
densities near the break radius would
help distinguish between these different possibilities.

\section*{Acknowledgments}

We thank David Thilker for sharing his HI data with us, 
Edouard Bernard for helpful advice about variable stars, 
Sebastian Hidalgo for useful discussions on the CMD fitting,
and the anonymous referee whose feedback helped improve the
content of this paper.
The optical image of M33's inner disc in Fig.\ \ref{fig:fields}
is provided courtesy of 
T.A.Rector (NRAO/AUI/NSF and NOAO/AURA/NSF) and M.Hanna (NOAO/AURA/NSF).
This paper is based on observations made with the NASA/ESA Hubble Space
Telescope, obtained at the Space Telescope
Science Institute, which is operated by the Association of
Universities for Research in Astronomy, Inc., under NASA contract
NAS 5-26555.  These observations are associated with program GO-9837.
This work has made use of the IAC-STAR Synthetic CMD computation code. 
IAC-STAR is supported and maintained by the computer division of the 
Instituto de Astrof\'{i}sica de Canarias.
This work has made use of the resources provided by 
the Edinburgh Compute and Data Facility (ECDF; 
http://www.ecdf.ed.ac.uk/). 
The ECDF is partially supported by 
the eDIKT initiative (http://www.edikt.org.uk).


\bibliographystyle{mn2e}

\bibliography{references}

\bsp



\label{lastpage}


\end{document}